\documentstyle[12pt,amssymb]{article} 
\renewcommand{\Bbb}{\bf}
\newcommand{\newsubsection}[1]{
\vspace{7mm}
\pagebreak[3]
\addtocounter{subsection}{1}
\addcontentsline{toc}{subsection}{\protect
\numberline{\arabic{section}.\arabic{subsection}}{#1}}
\noindent{\it \thesubsection.  #1}                 
\nopagebreak
\vspace{2mm}
\nopagebreak}
\newcommand{\newsection}[1]{
\vspace{15mm}
\pagebreak[3]
\addtocounter{section}{1}
\setcounter{equation}{0}
\setcounter{subsection}{0}
\setcounter{footnote}{0}
\addcontentsline{toc}{section}{\protect
\numberline{\arabic{section}}{{\rm #1}}}
\noindent{\bf \thesection.  #1}                 
\nopagebreak
\vspace{4mm}
\nopagebreak}
\renewcommand{\theequation}{\thesection.\arabic{equation}}

\newlength{\extraspace}
\newlength{\extraspaces}
\newcounter{dummy}
\newcommand{\be}{\begin{equation}
\addtolength{\abovedisplayskip}{\extraspaces}
\addtolength{\belowdisplayskip}{\extraspaces}
\addtolength{\abovedisplayshortskip}{\extraspace}
\addtolength{\belowdisplayshortskip}{\extraspace}}
\newcommand{\ee}{\end{equation}}
\newcommand{\ba}{\begin{eqnarray}
\addtolength{\abovedisplayskip}{\extraspaces}
\addtolength{\belowdisplayskip}{\extraspaces}
\addtolength{\abovedisplayshortskip}{\extraspace}
\addtolength{\belowdisplayshortskip}{\extraspace}}
\newcommand{\ea}{\end{eqnarray}}
\newcommand{\nonu}{\nonumber \\[2mm]}

\newcommand{\ban}{\begin{eqnarray*}
\addtolength{\abovedisplayskip}{\extraspaces}
\addtolength{\belowdisplayskip}{\extraspaces}
\addtolength{\abovedisplayshortskip}{\extraspace}
\addtolength{\belowdisplayshortskip}{\extraspace}}
\newcommand{\ean}{\end{eqnarray*}}
\newcommand{\baa}{                         
\addtocounter{equation}{1}
\setcounter{dummy}{\value{equation}}
\setcounter{equation}{0}
\renewcommand{\theequation}{\thesection.\arabic{dummy}\alph{equation}}
\begin{eqnarray}
\addtolength{\abovedisplayskip}{\extraspaces}
\addtolength{\belowdisplayskip}{\extraspaces}
\addtolength{\abovedisplayshortskip}{\extraspace}
\addtolength{\belowdisplayshortskip}{\extraspace}}
\newcommand{\eaa}{                                       
\end{eqnarray}
\setcounter{equation}{\value{dummy}}
\renewcommand{\theequation}{\thesection.\arabic{equation}}}

\input epsf
\newcounter{fignum}
\renewcommand{\d}{{\partial}}

\newcommand{\ie}{{\it i.e.}}
\newcommand{\etc}{{\it etc.\ }}

\newcommand{\eg}{{\it e.g.\ }}

\renewcommand{\l}{\langle}

\newcommand{\tr}{{\rm tr}}

\newcommand{\half}{{\textstyle{1\over 2}}}

\newcommand{\Z}{{\Bbb Z}}
\newcommand{\R}{{\Bbb R}}

\newcommand{\C}{{\Bbb C}}

\renewcommand{\H}{{\Bbb H}}

\newcommand{\ra}{\rightarrow}

\newcommand{\oprra}{\mathop{\longrightarrow}}

\newcommand{\inv}{^{-1}}

\newcommand{\Tr}{{\rm Tr}\,}

\newcommand{\th}{^{\mit th}}

\newcommand{\cC}{{\cal C}}

\newcommand{\cE}{{\cal E}}
\newcommand{\cB}{{\cal B}}

\newcommand{\cM}{{\cal M}}
\newcommand{\cF}{{\cal F}}
\newcommand{\cN}{{\cal N}}

\newcommand{\End}{{\rm End}}

\newcommand{\ext}{{\raisebox{.2ex}{$\textstyle \bigwedge\!$}}}

\def\a{\alpha} 
\def\b{\beta} 
\def\g{\gamma} 
\def\G{\Gamma}

\def\th{\theta}

\def\l{\lambda} 
 
\def\m{\mu}
\def\n{\nu}
 
\def\s{\sigma} 

\def\f{\phi} 
 
\def\w{\omega}

\def\z{\zeta}
\def\<{\langle}
\def\>{\rangle}

\newfont{\gothic}{eufm10 scaled\magstep1}

\newcommand{\Sym}{{S}}
\renewcommand{\bar}{\overline}

\newcommand{\zz}{{\hbox{\boldmath $\z$}}}
\newcommand{\bb}{{\hbox{\boldmath $b$}}}
\newcommand{\ii}{{\hbox{\boldmath $i$}}}
\newcommand{\jj}{{\hbox{\boldmath $j$}}}
\newcommand{\kk}{{\hbox{\boldmath $k$}}}
\newcommand{\AR}{{\cC}}
\newcommand{\BR}{{\widetilde{B}}}

\begin{document}

\addtolength{\baselineskip}{.5mm}
\thispagestyle{empty}
\begin{flushright}
October 1998\\
{\sc hep-th/9810210}\\
{\sc utfa-98/26}\\
{\sc spin-98/4}
\end{flushright}

\vspace{8mm}

\begin{center}
{\Large \sc Instanton Strings and HyperK\"ahler Geometry}
\\[25mm] {Robbert Dijkgraaf\footnote{\tt rhd@wins.uva.nl}}\\[4mm]
{\it Departments of  Mathematics and Physics\\
University of Amsterdam, 1018 TV Amsterdam}\\[2mm]
{\cal \&}\\[2mm]
{\it Spinoza Institute\\
University of Utrecht, 3508 TD Utrecht}
\\[2cm] 
{\sc Abstract}
\end{center}

\noindent We discuss two-dimensional sigma models on moduli spaces of
instantons on $K3$ surfaces. These $\cN=(4,4)$ superconformal field
theories describe the near-horizon dynamics of the D1-D5-brane system
and are dual to string theory on $AdS_3$. We derive a precise map
relating the moduli of the $K3$ type IIB string compactification to the
moduli of these conformal field theories and the corresponding classical
hyperk\"ahler geometry. We conclude that in the absence of background
gauge fields, the metric on the instanton moduli spaces degenerates
exactly to the orbifold symmetric product of $K3$. Turning on a
self-dual NS $B$-field deforms this symmetric product to a manifold that
is diffeomorphic to the Hilbert scheme. We also comment on the
mathematical applications of string duality to the global issues of
deformations of hyperk\"ahler manifolds.

\vfill

\newpage

\newsection{Introduction}

Recently in string theory there has been great interest in a particular
class of two-dimensional conformal field theories with $\cN=(4,4)$
supersymmetry and central charge $c=6k$ with $k$ a positive integer.
These conformal field theories arise by considering a certain low-energy
limit of bound states of D5-branes and D1-branes in type IIB string
theory and in various other dual incarnations. They have been
instrumental in the microscopic description of the quantum states of
five-dimensional black holes as in the ground-breaking computation of
Strominger and Vafa \cite{strominger-vafa}.

The same sigma models have appeared in the quantization of the
six-dimensional world-volume theory of the NS fivebrane, that can be
used to describe black holes in matrix theory
\cite{5brane,5d,abkss,witten-higgs,abs,maxmic}. In this context one
sometimes refers to these models as second-quantized microstrings,
little strings or instanton strings. 

More recently, in the work of Maldacena on dualities between space-time
conformal field theories and supergravity theories on anti-de Sitter
spaces, these models were identified with the dual description of
six-dimensional string theory on $AdS^3 \times S^3$ \cite{maldacena}.
The holographic correspondence between these $AdS$ quantum gravity
theories and two-dimensional conformal field theories with their rich
algebraic structure, seems to be a fertile proving ground for various
conceptional issues in gravity and string theory, see \eg the recent
papers \cite{strominger,maldacena-strominger,martinec,deboer,bdhm} and
references therein to the already extensive literature. Finally, there
is a fascinating relation between the world-sheet CFT of a string moving
in this $AdS^3$ background and the space-time $c=6k$ CFT that we study
in this paper \cite{gks}. 

This particular class of $\cN=(4,4)$ SCFT with central charge $6k$ can
be identified with sigma models on moduli spaces of instantons on $K3$
or $T^4$ \cite{strominger-vafa}. With the right characteristic classes
and compactification, these moduli spaces are smooth, compact $4k$
dimensional hyperk\"ahler manifolds. Independent of the developments in
string theory there has been recently mathematical interest in these
higher-dimensional compact hyperk\"ahler (or holomorphic symplectic)
manifolds and their deformation spaces, following the important work of
Mukai on moduli spaces of sheaves on abelian and $K3$ surfaces
\cite{mukai,mukai-vector,mukai-map}. See for example \cite{huybrechts}
for a good review of the recent developments.

In this paper we will see how the insights from string theory and from
hyperk\"ahler geometry combine nicely and allow us to make a precise
identification between the moduli of the string compactification and
those of the (space-time) CFT.  Roughly we find two set of canonical
identifications: 

(1) between the moduli of a special (attractor) family of world-sheet
$K3$ sigma model and the classical hyperk\"ahler geometry of the
instanton moduli space;

(2) between the moduli of a special family of non-perturbative IIB
string theory compactification on $K3$ and the $(4,4)$ superconformal
field theory on the instanton moduli space.

We will see in detail how both the metric and $B$-field on $K3$ are used
to determine the metric on the compactified instanton moduli space. The
RR fields then encode the $B$-field of the sigma model. Furthermore,
non-perturbative $U$-dualities of the IIB string compactification
translate into $T$-dualities of the hyperk\"ahler sigma model.

The outline of this paper is as follows. In \S2 we introduce the
D5-brane world-volume theory and its relation to a sigma model on the
instanton moduli space. In \S3 we will describe in detail the moduli of
$K3$ compactifications in string theory. Then in \S4 we will use this
explicit description to determine the value of these moduli on D-brane
bound states that minimalize the BPS mass through the so-called
attractor mechanism. We will treat in detail the examples of the
D0-D4-brane system and D2-branes in IIA theory (or equivalently D1-D5
and D3 in IIB theory). In \S5 we will review some of discussion of
hyperk\"ahler manifolds and their deformation spaces in the mathematical
literature. Then in \S6 we will use these mathematical facts to give an
explicit map between the $K3$ moduli and the hyperk\"ahler structures on
the instanton moduli space, a map that we will then extend in \S7 to
string theory moduli and the moduli of the $\cN=(4,4)$ SCFT. We will end
with some comments on global issues of these moduli spaces.

\newsection{D5-branes and  instanton strings}

\vspace{-4mm}

\newsubsection{Some mathematical preliminaries}

In this paper we will be studying sigma models with target space the
moduli space $\cM$ of instantons on a four-manifold $X$. So,
mathematically speaking, we are interested in the quantum cohomology or
more generally the Gromov-Witten invariants of $\cM$. The classical
cohomology of $\cM$ is essentially equivalent to the Donaldson
invariants of $X$, so the sigma model on $\cM$ defines a ``stringy''
generalization of these invariants --- we will refer to them as
instanton strings. However, we should hasten to add that for the
particular case of a $K3$ or $T^4$ manifold, on which we will mostly
concentrate in this paper, the moduli space is a hyperk\"ahler manifold,
and the quantum cohomology reduces to the ordinary cohomology. So one
should turn in this case to more refined string theory invariants such
as the elliptic genus \cite{strominger-vafa,ell-genus}.

The quantum cohomology of the moduli space of vector bundles on a
Riemann surface is well-known to play an important role in the study of
instantons on 4-manifolds that can be written as the product of two
Riemann surfaces or, more generally fibrations of this form
\cite{ds,bjsv}. If $X$ is of the form $\Sigma \times \Sigma'$, in the
adiabatic limit where the volume of $\Sigma'$ is very small compared to
$\Sigma$, instantons on $X$ reduce to holomorphic maps of $\Sigma$ into
the moduli space of holomorphic vector bundles on $\Sigma'$. In a
similar spirit the quantum cohomology of instanton moduli spaces on a
four-manifold $X$ should be related to invariants of Yang-Mills gauge
theories on sixfolds (or complex threefolds) of the form $Y=\Sigma
\times X$. 

Now, naively, non-abelian quantum gauge theories do not make sense
beyond space-time dimension four because they become strongly
interacting at short distances, which ruins the renormalizability of the
theory.\footnote{Of course, one of the important lessons we have learned
from the success of Seiberg-Witten theory is that in topological
applications it can suffice to work with an {\it effective} quantum
field theory, that only makes sense up to a certain cut-off distance
scale.} Six-dimensional supersymmetric Yang-Mills can be made sense of
as a quantum system in string theory, namely as the world-volume theory
of multiple Dirichlet fivebranes \cite{polchinski}, with strings
providing the natural regulator at short distances \cite{nati}. At low
energies the world-volume theory of $N$ of these Dirichlet fivebranes
reduces to a six-dimensional $U(N)$ super-Yang-Mills theory. Quantum
fivebranes are one of the most mysterious but also mathematically
richest objects in string theory, see \eg \cite{witten-M5,icm}. Via
$T$-dualities they are closely related to the study of moduli spaces of
stable holomorphic vector bundles on $Y$, a problem that has very
important applications in string theory and that has received some
recent mathematical impetus \cite{dt,thomas}.

\newsubsection{The D5-brane action}

Ignoring fermions and scalar fields, the leading part of the D5-brane
world-volume action on a six-manifold $Y$ embedded in ten-dimensional
space-time (we will assume with a trivial normal bundle) is of the form 
\be
S = \int_Y {1\over g_s} \Tr \cF \wedge * \cF
+ \AR \wedge v'(\cF).
\label{S}
\ee
Let us explain the various objects in this Lagrangian: $g_s \in \R_+$
is the type IIB string coupling. The covariant field strength $\cF$ is
defined as \cite{witten-D}
\be
\cF = F - 2\pi i B,
\label{F}
\ee
with $F$ the usual curvature of the $U(N)$ connection on the rank $N$
vector bundle $\cE$ over $Y$, and $B$ the background NS tensor field, a
harmonic 2-form on $Y$ (actually, the pull-back to $Y$ of the
ten-dimensional space-time $B$-field.) Note that $B$ is a singlet under
$U(N)$, so it only couples to $\Tr F$. 

The background RR gauge field $\AR$ is the pull-back to $Y$ of
an arbitrary harmonic form of even degree in space-time
and it can be decomposed as
\be
\AR = \theta + \BR + G
\ee
with $\theta$ a scalar, $\BR$ the RR 2-form field and $G$ a 4-form
with a field strength that satisfies the self-duality constraint 
in ten dimensions, $dG=*dG$. These RR fields couple to
the 5-brane through the generalized Mukai vector $v'$ given by
\cite{li,douglas,ghm,harvey-moore-2}
\be
v' = \Tr \exp\left({i\cF\over 2\pi}\right) \wedge \widehat{A}(Y)^{1/2}.
\label{v}
\ee
Here the first term is a generalization of the usual Chern character
$ch(\cE)$ including the NS $B$-field. The expression $\widehat{A}(Y)$ 
that figures in the second term is the so-called A-roof genus of the 
manifold $Y$. It appears in the index theorem of the Dirac operator
and can be expressed as a particular combination of Pontryagin classes 
of $Y$. For a Calabi-Yau threefold (or twofold) 
the $\widehat A$ genus equals the Todd genus 
\be
Td(Y) = 1 + {c_2\over 12},
\ee
and we can write the RR charge vector as
\be
v' = \Tr\exp\left({iF\over 2\pi} + B +  {c_2\over24}\right).
\ee

Writing out the coupling of the RR gauge fields (on a flat space and
ignoring the NS $B$-field) we obtain schematically a combination of
the form
\be
\int \th\; \Tr F^3 + \BR \wedge \Tr F^2 + G \wedge \Tr F
\ee
{}From this we see that Yang-Mills instantons, which contribute to the
second Chern character $\Tr F^2$, carry charge with respect to
$\BR$ and can therefore be interpreted as bound states with D1-strings.
Similarly D3-brane bound states couple to the four-form $C$ and are
carried by gauge field configurations with a non-trivial first Chern
class or magnetic flux $\Tr F$.

If we consider a 6-manifold of topology $\Sigma \times X$ (or a local
$X$-fibration) in the adiabatic limit where the volume of the surface
$\Sigma$ is much larger than the four-manifold $X$, the SYM theory
reduces to a sigma model with target space the moduli space of
anti-self-dual connections $F_+=0$ on $X$. (Here we assumed that the
instanton number is positive, otherwise we should consider self-dual
connections or bound states with anti-D1-branes.) If we include a background {\it
self-dual} harmonic $B$-field, the abelian part of the ASD equation 
is deformed to
\be
F_+ = 2\pi i B.
\ee

The various terms in the D5-brane action (\ref{S}) now acquire a sigma
model interpretation. The kinetic term gives a natural metric $g$ on the
instanton moduli space which is equivalent to the usual $L^2$-metric on
adjoint-valued one-forms on $X$, defined for $\a,\b \in T\cM$ as
\be
g(\a,\b) = {1\over g_s} \int_X \Tr (\a\wedge *\b),
\ee
where the tangent vectors can be chosen to satisfy
$(D\a)_+=D^*\a=0$. From the normalization of the kinetic term we see
that with this metric the volume of $\cM$ scales with a power of
$1/g_s$ and goes to infinity in the weak coupling limit $g_s\to 0$.

If the four-manifold $X$ is hyperk\"ahler (HK) with hyperk\"ahler
forms $\vec\w=(\w_1,\w_2,\w_3)$, the instanton moduli space itself is
a hyperk\"ahler manifold. In fact, it can be considered as an
infinite-dimensional HK quotient with HK moment map
\be
\vec\m = F \wedge \vec\w.
\ee
The condition $\m=0$ enforces the ASD equation.  With $\cal G$ the
infinite-dimensional gauge group, the moduli space of instantons $\cM$
can therefore be written as
\be
\cM = \m\inv(0)/{\cal G}
\ee
and this guarantees that it is hyperk\"ahler \cite{hitchin}. From this
point of view turning on a self-dual $B$-field on $X$ corresponds to
a change in the value of the hyperk\"ahler moment map. In some
cases this deformation can resolve certain singularities \cite{nakajima}.

\newsubsection{Topological observables and the Donaldson-Mukai map}

The coupling of the RR background gauge fields to the D5-brane gauge
theory produces natural cohomology classes on the instanton moduli
space. In particular, it induces two kinds of natural closed 2-forms on
$\cM$ that acquire an interpretation as $B$-fields in the sigma model.
These two-forms are well-known from the polynomial Donaldson invariants
\cite{donaldson-kronheimer} and Witten's topological field theory
realization of these invariants \cite{tft}. It is remarkable that in this
case we can naturally identify the deformation parameters of quantum
cohomology with the standard variables of the four-manifold
invariants.

First, for any harmonic two-form on $X$ we have the
well-known two-dimensional cohomology clas  on $\cM$ defined by
Donaldson. With $\cB_I$ ($I=1,\ldots,b_2$) a basis of $H^2(X,\Z)$, 
these classes take the local form
\be
\cB_I(\a,\b) = \int_X B \wedge \Tr(\a \wedge \b).
\label{BR}
\ee
There is a second natural two-form, namely
\be
\cB_0(\a,\b) = \int_X \Tr( F \wedge \a \wedge \b).
\label{BF}
\ee
In this way we obtain a natural set of $b_2+1$ two-forms on the moduli
space $\cM$. We will see later that for $K3$ surfaces these forms form
a basis for $H^2$ of the instanton moduli space.

Note that in the case of stable {\it holomorphic} vector bundles on a
complex surface $X$, the holomorphic tangent vectors $\a,\b$ are
elements of $H^{0,1}(X,\End\,\cE)$ and the curvature $F$ is of type
(1,1). Now we also have a correspondence of Hodge structures: a (2,0)
form $B_I$ gives a (2,0) form $\cB_I$ on moduli space \etc The
additional two-form $\cB_0$ is always of type (1,1).

In topological field theory terms these two-form are operators obtained
through the so-called descent formalism \cite{tft}. The forms $\cB_I$
are the two-form descendents of the BRST invariant operator $\Tr \f^2$ ;
the second form $\cB_0$ is similarly a four-form descendent of
$\Tr\f^3$, an operator that only appears as an independent field in
$U(N)$ gauge theory with $N>2$.

Mathematically, these two-forms are produced using (generalizations of)
Donaldson's $\mu$ map that associates a cohomology class on the moduli
space to a cohomology class on the four-manifold
\be
\mu_2 : \ H^k(X) \ra H^k(\cM),
\ee
and which is defined as follows. One considers the universal bundle $\widehat\cE$ on the product 
$X \times \cM$. This bundle has a curvature $\widehat F$ and
a second Chern character $\Tr \widehat F
\wedge \widehat F$. For any form $\AR \in H^*(X)$ one now defines a 
cohomology class on $\cM$ by multiplying with the Chern class and
then integrating over the fiber $X$
\be
\mu_2(\AR) = \int_X \AR \wedge \Tr(\widehat F \wedge \widehat F).
\ee
With this definition we have 
\be
\cB_I = \mu_2(B_I). 
\ee
Similarly one can define a map $\mu_3$ from $H^k(X)$ to $H^{k+2}(\cM)$ 
starting from $\Tr \widehat{F}^3$ and with this map we have 
$\cB_0 = \mu_3(1)$.

Actually, from the precise coupling (\ref{S}) to the generalized Mukai
vector $v'$ we see that the relevant map induced by the D5-brane
is the generalized Mukai map \cite{mukai-map} 
\be
\mu:\ H^*(X) \ra H^*(\cM)
\ee
which does not preserves degrees and which is roughly defined as
\be
\mu(\AR) = \int_X \AR \wedge \hat v,\qquad \hat v=e^B \, ch(\hat\cE)
\, \widehat{A}^{1/2}(X\times \cM).
\label{mu}
\ee
We want to stress that this coupling through the Mukai vector gives a
natural map from the harmonic RR background fields to the cohomology of
the instanton moduli space. We will return to the properties and
interpretation of this important map.

\newsection{String theory on $K3$}

Although almost everything discussed in this paper pertains to string
compactifications on both four-tori and $K3$ manifolds, we will mostly
concentrate on the latter. We will first recall some facts about
string theory on $K3$. An excellent review of the various applications
of $K3$ surfaces in string theory is given in \cite{aspinwall}. 

\newsubsection{Classical and CFT moduli of $K3$}

Let $X$ be a $K3$ surface. Recall that $H^2(X,\Z)$ equipped with the
intersection product is isomorphic to $\G^{3,19}$, the unique even,
self-dual lattice of signature $\bigl((+)^3,(-)^{19}\bigr)$. By the
global Torelli theorem of $K3$ surfaces and Yau's theorem, a
hyperk\"ahler structure on a marked $K3$ surface is uniquely
determined by the positive 3-plane $U\subset H^2(X,\R)$ spanned by the
periods of the three HK forms
\be 
U = \langle \w_1,\w_2,\w_3\rangle
\ee 
together with the volume $V\in \R_+$. With this metric $U \cong
H^2_+(X,\R)$, the space of self-dual two-forms on $X$.  It will turn
out to be convenient to normalize the HK forms $\vec\w$
(unconventionally) such that $\half \w_a \wedge \w_b = \delta_{ab}
V/3.$

Since every positive three-plane $U$ can appear as the image under the
period map of a HK structure, the moduli space of HK metrics of fixed
volume (including singular orbifold metrics, where $-2$ curves are
contracted and symmetry enhancing singularities of $ADE$-type occur)
is given by the Grassmannian
\be
{O(3,19) \over O(3) \times O(19)}. 
\label{3,19}
\ee
The corresponding moduli space for unmarked $K3$ surfaces is obtained
by further dividing by $O^+(\G^{3,19})$, the component group of the
group of diffeomorphism.

If we choose a compatible complex structure, the positive 3-plane $U$
decomposed as the orthogonal som $P\perp L$, with $P$ the 2-plane
spanned by the real and imaginary components of the holomorphic
$(2,0)$ form, say $\eta=\w_2+i\w_3$, and $L$ the (oriented) line
generated by the real $(1,1)$ K\"ahler form, say $\w_1$.  The form
$\eta$ spans $H^{2,0}(X)$ and its periods take value in the period
domain
\be
\eta^2 =0,\qquad (\eta + \bar\eta)^2>0.
\ee
The global Torelli theorem guarantees that the periods of $\eta$
uniquely determine the complex structure on $X$.  For a given
hyperk\"ahler structure there is a $S^2$ worth of inequivalent
compatible complex structures, corresponding to the choices of
$P\subset U$.

In conformal field theory one further adds the NS 2-form background
field $B\in H^2(X,\R/\Z)$. So locally the moduli space of $K3$ CFT's
is
\be
{O(3,19)\over O(3)\times O(19)} \times \R^{3,19} \times \R_+ .
\ee
Supersymmetry arguments tell us that this component of the moduli
space of $\cN=(4,4)$ superconformal field theories with $c=6$ is
locally given by the symmetric space \cite{seiberg}
\be
{O(4,20) \over O(4) \times O(20) }.
\label{4,20}
\ee
A precise analysis of the quantum symmetries shows that moduli space
of inequivalent $K3$ sigma model superconformal field theories is
obtained as a further quotient by the $T$-duality group $O(\G^{4,20}) =
O(4,20,\Z)$, with $\G^{4,20}$ the even, self-dual lattice of signature
$(4,20)$ \cite{aspinwall-morrison}.

\newsubsection{The Mukai vector}

The moduli space (\ref{4,20}) has an obvious interpretation as the
Grassmannian parametrizing positive 4-planes $W \subset \R^{4,20}$. This
representation follows most naturally by considering a compactification
of the type IIA (or IIB) string on $X$. Since the RR fields in IIA
theory have even-dimensional curvatures, the lattice of RR charges
carried by D-brane bound states (that are point-like in the
uncompactified six dimensions) is given by $H^*(X,\Z)$. Equivalently,
instead of looking at point-like objects, one could look at
six-dimensional strings that are obtained by wrapping odd D-branes in
type IIB theory, as we will later. 

A D-brane state with charge $v\in H^*(X,\Z)$ can be described in terms 
of a vector bundle, or more generally a coherent sheave $\cE$ over $X$ 
with Mukai vector
\be
v=ch(\cE)Td(X)^{1/2}=(r,c_1,r+ch_2).  
\ee
Here the Chern character is defined as $ch(\cE)=\tr \exp iF/2\pi$ and
$r$ is the rank of $\cE$, $ch_2 = -c_2 + \half c_1^2$. 
Note that here we did not use the definition (\ref{v})
of the generalized Mukai vector $v'$, which includes the effect of the
$B$-field. In fact the two are simply related as
\be
v' = e^B \wedge v.
\ee
The point is that because of the $B$-field contribution the generalized
Mukai vector is generically no longer an integer cohomology class, and 
we like to fix the charge lattice once and for all and identify it with 
$H^*(X,\Z)$. We will instead account for the effect of the $B$-field in terms of the moduli of the
$K3$ sigma model. 

It is natural to give $H^*(X,\Z)$ the Mukai intersection product,
defined as \cite{mukai-vector}
\be
v\cdot v = \int_X (v^2 \wedge v^2 - 2 v^0 \wedge v^4),
\qquad v=(v^0,v^2,v^4),\ v^i\in H^i(X).
\ee
We will always identify $H^*(X,\Z)\cong \G^{4,20}$ with Mukai's
quadratic form. With this definition the moduli space $\cM_v$ of simple
sheaves with Mukai vector $v$ has complex dimension 
\be
\dim \cM_v = 2 + v\cdot v.
\ee

\newsubsection{A quaternionic formula}

The map that associates a real 4-plane $W \subset H^*(X,\R)$ to the
three HK forms $\vec\w$ and the $B$-field can be elegantly formulated
as follows. Combine the four 2-forms into a single quaternionic 2-form
\be
\bb = B + \w_1 \ii + \w_2 \jj + \w_3 \kk.
\ee
The vector $\bb$ should be considered to take value in the quaterionic
domain
\be
\left\{ \bb \in \H^{3,19} \mid \Im \bb >0 \right\}.
\ee
With this notation the 4-plane $W$ can be considered as a quaternionic
line in $H^*(X,\R) \otimes \H$, given in terms of $\bb$ as
\be
W = \langle \exp\bb \rangle =\langle 1 + \bb + \half \bb \wedge \bb \rangle.
\label{quaternion}
\ee
Note that $\exp\bb$ is a quaternionic null vector in the Mukai inner
product.  In components the plane $W$ is spanned by the
vectors\footnote{Here we used the normalization $\half \vec\w\cdot
\vec\w=V$. Note a small discrepancy with the formula of
\cite{aspinwall,moore-aa} regarding the precise contribution of the
$B$-field.}
\be
(0,\vec\w,B\wedge \vec\w),\ \ 
(1,B,\half B\wedge B - V).
\ee
Note that the volume $V$ is measured in terms of the string length
$\sqrt{\a'}$, so the classical limit $\a'\to 0$ implies $V\to \infty$.

There are several remarks that might help to understand formula
(\ref{quaternion}) better. First, note that $\bb$ can be written in
terms of the complexified K\"ahler form $\w=B + i\w_1$ and the
holomorphic $(2,0)$ form $\eta=\w_2 + i \w_3$ (that determines the
complex structure) as
\be
{\bb} = \w + \eta {\jj}.
\ee
In this notation we see that a mirror symmetry map that interchanges
the complex and K\"ahler structure acts as a simple quaternionic
rotation by multiplying by $\jj$.

Secondly, from the action (\ref{S}) we see that turning on the
$B$-field, shifts $F \ra F - 2\pi i B$ in the expression of the charge
vector
\be
v \ra e^B \wedge v = v',
\ee
which is in line with our formula. 

Thirdly, the representation (\ref{quaternion}) can be seen as a
quaternionic generalization of an analogous complex representation for
the hermitean symmetric domain
\cite{borcherds, harvey-moore}
\be
{O(2,2+s) \over O(2) \times O(2+s)} \cong  
\left\{ y\in \C^{1,1+s} \mid \Im y>0,\ y^{1,0}>0 \right\}.
\ee
To prove this isomorphism one associates to such a vector $y$ the
complex null vector
\be
z=(y,1,-\half y^2) \in \C^{2,2+s}.
\ee
(Here we note the similarity with (\ref{quaternion}).)  The
corresponding positive two-plane in $\R^{2,2+s}$ is then spanned by
$\Re z, \Im z$.

Finally, in the case of a four-torus $T=\R^4/2\pi L$ with lattice $
L\cong \Z^4$, we have a similar formula that can be understood along
more familiar lines.  In the torus case the $T$-duality group
$SO(4,4,\Z)$ is usually considered to act in the ${\bf 8}_v$ vector
representation on the Narain lattice
\be
L \oplus L^* \cong \G^{4,4},
\ee
equipped with the standard quadratic form
\be
v\cdot v=2 \langle w, k\rangle ,\qquad v=w \oplus k.
\ee
Picking a basis $e_\m$ of $L$ and dual basis $e^\m$ of $L^*$, the data of 
a flat metric  $G_{\m\n}$ and constant $B$-field $B_{\m\n}$ determine a 
positive four-plane in $\R^{4,4}$ spanned by the vectors
\be
e_\m + X_{\m\n} e^\n,
\qquad 
X_{\m\n}= G_{\m\n} + B_{\m\n}.
\ee
The $T$-duality group acts by fractional linear transformations on the
$4\times 4$ matrix $X_{\m\n}$.

However, by triality the group $SO(4,4,\Z)$, or more properly its cover
$Spin(4,4,\Z)$, also acts on the spinor representation ${\bf 8}_s$,
that can be  canonically identified with IIA string theory
RR charge lattice
\be
H^{even}(T,\Z)\cong \ext^{even} L^* \cong \G^{4,4}.
\ee
Similarly for the type IIB string we have an action in the ${\bf 8}_c$
representation on the lattice $H^{odd}(T,\Z)$. In this spinor
representation the corresponding positive 4-pane $W$ in
$\ext^{even}[e^\m]\cong \R^{4,4}$ is determined as the positive
eigenspace of the generalized Hodge $*$-operator
\be
* = \g_1\wedge\g_2\wedge\g_3\wedge\g_4,
\ee
written in terms of the Clifford algebra generators
\be
\g_\m = {\d\over\d e^\m} + X_{mn} e^\m.
\ee
One now easily checks that $W$ can be represented in the above form
(\ref{quaternion}) with $\w_a$ a basis of self-dual 2-forms. For
example, in the above formula it is clear that a shift of $B$ is
implemented by conjugating $\g_\m \ra e^{-B} \g_\m e^B$.

\newsection{D-branes and attractive $K3$'s}

We will now apply this explicit description of the $K3$ conformal field
theory moduli to D-brane bound states. To this end let us consider a
general D-brane state labeled with a Mukai vector
$v\in \G^{4,20}$, $v^2>0$ and $v$ primitive. We will assume that
\be
v^2 = 2k-2,\qquad k >1.
\ee
Such a vector defines an orthocomplementary lattice $v^\perp \subset
\G^{4,20}$ of signature $(3,20)$.
The lattice is even but not self-dual (unimodular).

One can use some well-known results in lattice theory\footnote{I wish to
thank E.~Looijenga for a very helpful discussion about this point.}
\cite{lp,nikulin} to prove that any two D-brane charge vectors of equal
length are related by a $T$-duality transformation in
$O(\G^{4,20})=O(4,20,\Z)$. Indeed, the orthogonal group acts
transitively on the set of primitive vectors of a given length in a
lattice, if the lattice contains at least 3 hyperbolic lattices
$H\cong\G^{1,1}$. Since we have the isomorphism \be \G^{4,20} \cong 4 H
\oplus (-2 E_8), \ee with $E_8$ the root lattice of the corresponding
Lie algebra, this is indeed the case. This implies that all lattices
$v^\perp$ with fixed $v^2=2k-2$ are isomorphic. We sometimes write this
lattice as $\G^{3,20}_k$ or just simply $\G^{3,20}$, but note that the
integer quadratic form of signature $(3,20)$ depends on the value of
$k$.

Furthermore, we can always find a primitive  vector $u \in v^\perp$ 
satisfying
\be
u\cdot v = 0,\qquad u^2=-v^2=2-2k,
\ee
such that the lattice decomposes as
\be
\G^{3,20}_k \cong \G^{3,19} \oplus \Z\cdot u.
\ee
Again all choices of such a vector $u$ are related by a stabilizing
$T$-duality transformation in $O(v^\perp) = O(3,20,\Z)$. To prove this
one has to study primitive embeddings of the hyperbolic lattice
$\langle u,v\rangle$ in $\G^{4,20}$.  Since this lattice has rank two,
such an embedding is unique if the target lattice contains at least
{\it four} hyperbolic factors, which it indeed does (or equivalently,
$\G^{3,20}_k$ should contain at least three hyperbolic factors.)

We will now look at the $K3$ sigma model moduli that minimalize the BPS
mass of the corresponding D-brane state. These are the fixed point
values of the near-horizon moduli as described by the so-called
attractor mechanism \cite{fks}, see also the extensive analysis of
\cite{moore-aa} and references therein.\footnote{Note however that the
attractive $K3$ surfaces described here differ from the $K3$ surfaces
that appear in $K3\times T^2$ compactifications and that are described
in detail in \cite{moore-aa}. The latter have much more subtle
arithmetic properties than the simple families described here. In
particular here we have a unique $T$-equivalency class for given
``discriminant'' $-v^2$.} In the supergravity limit these are the values
of the scalar fields on the D-brane horizon, see \eg \cite{aafl}. In the
quantum description of the D-brane degrees of freedom these are the
relevant values of the $K3$ moduli.

Given a positive 4-plane or polarization $W\subset \R^{4,20}$, 
the charge vector $v$ decomposes as
\be
v = v_L^{4,0} + v_R^{0,20}\!\!,\qquad v^2 = v_L^2 - v_R^2.
\ee
The BPS mass formula is given by 
\be
m^2 = v_L^2=v^2 - v_R^2,
\ee
and this is clearly minimalized if $v_R=0$, \ie\ if the charge vector
$v$ is contained in the 4-plane $W$. Writing $W$ as $U \perp \R v$, we
see that the restricted moduli space that preserves the minimal BPS
mass is locally given by the Grassmannian of positive 3-planes $U$ in
the complement $v^\perp\otimes \R \cong \R^{3,20},$ and can therefore
be written as
\be
{O(3,20) \over O(3) \times O(20)}.
\label{3,20}
\ee
Furthermore, we have the remnant of the $T$-duality group $O(v^\perp)
\cong O(3,20,\Z)$, that preserves the charge vector $v$ and is therefore
an automorphism of the D-brane state. The quantum restricted moduli 
space is obtained by further quotienting by this $T$-duality subgroup.

It might be interesting to consider these attractor moduli in a few
explicit special cases, see also the discussion in
\cite{harvey-moore-2}. We use the notation of IIA theory but the result
equally well applies to the IIB theory.

\newsubsection{D0-D4-branes}

Let us first consider only D0 and D4 branes, no D2-branes \cite{vafa}.
Here we pick a Mukai vector of type $v=(r,0,-p)$, with $r\cdot p > 0$
and $r,p$ relative prime. In the gauge theory this is the case of a
vector bundle of rank $r$, $c_1=0$ and instanton number $c_2=r+p.$

The BPS conditions now read (with $\half \vec\w\cdot \vec\w =V$, and
$\vec\zeta\in\R^3$ to be determined)
\be
\vec\zeta \cdot \vec\w + r\,B = 0,
\ee
\be
\vec\zeta \cdot \vec\w \wedge B + r\, (\half B\wedge B - V) = -p.
\ee
These equations can be simply solved by first choosing 
$B= - \vec\zeta \cdot \vec\w/r$, so that for the given 
(unrestricted) HK structure $\vec\w$ the $B$-field should 
be self-dual, 
\be
B \in H^2_+(X)\cong \R^3.
\ee
We have already commented on the interpretation of adding a SD $B$-field
in terms of shifting the value of HK moment map in \S2. 
Secondly, the overall volume of the $K3$ (in units of $\a'$) should 
be fixed to be
\be
V   = p/r - \half B^2.
\ee
So, for small charges an attractive $K3$ is typically of the size 
of the string, and the sigma model is strongly coupled.
Note that since $B$ is self-dual, $B^2$ is positive, as is $V$, so
that consequently we have a bound on the $B$-field of the form $B^2
\leq 2p/r$. Combining the ingredients we see that in this case the
restricted moduli space (\ref{3,20}) decomposes locally as
\be
{O(3,19)\over O(3) \times O(19)} \times \R^3.
\ee

\newsubsection{D2-branes}

The example of D2-branes has been studied among others in \cite{bsv}.
Here the Mukai vector is given as $v=(0,q,0)$ with $q\in H^2(X,\Z)$
primitive and satisfying $q \cdot q >0$. In gauge theory terms we are
considering a sheaf localized on a complex curve with $c_1=q$ and
$ch_2=0$. Now the BPS conditions read
\be
\vec\zeta \cdot \vec\w  =  q,\qquad
\vec\zeta \cdot \vec\w \wedge B  = 0.
\ee
The first equation implies that the HK structure must be chosen such that
\be
q \in H^2_+(X).
\ee
This in turn requires that  $H^2_+(X) \cap H^2(X,\Z)$ should be at least
one-dimensional. The total volume of the $K3$ surface is in this 
case unrestricted.  Furthermore, according to the second equation,
the $B$-field should satisfy
\be
B \cdot q = 0,
\ee
so that $B\in q^\perp \otimes \R \cong \R^{2,19}$.
For D2-branes the moduli space (\ref{3,20}) therefore
decomposes in the familiar way as
\be
{O(2,19)\over O(2)\times O(19)} \times \R^{2,19} \times \R_+ .
\ee

Equivalently, there must exist a compatible complex structure on $X$ such
that $q \in H^{1,1}(X)$ and positive. Therefore the Picard lattice
should be at least one-dimensional. That in turn implies that $q$ is
Poincar\'e dual to a homology class $[C]$ of a holomorphic curve $C$ in
$X$ of genus $g=1 + \half q \cdot q \geq 2$. That is, the $K3$ moduli
are such that $q$ can be realized as a supersymmetric two-cycle, a
well-known result \cite{bsv,yz}. Furthermore the $B$-field flux
through this curve is required to vanish,
\be
\int_C B=0.
\ee

\newsubsection{General case}

After these two examples it is not difficult to write down the formulas for 
the fixed point moduli for the general D-brane bound state system with 
$v=(r,q,p)$ primitive,  $v^2 = 2 rp + q^2 >0$, since such a vector can be
obtained from the previous two examples by simply shifting the $B$-field.

For $r=0$ we have D0 and D2-branes. Since we obtain this case from example
4.2 by shifting $B$ by a 2-form $B_0$ satisfying $B_0\cdot q=p$,
the only difference is now that the total flux of the $B$-field satisfies
\be
\int_C B = p.
\ee

If $r\neq 0$ we can write $v=e^B \wedge v'$ with
$v'=(r,0,p')=(r,0,v^2/2r)$ and $B=q/r$. (We should not worry about $p'$
not being integer.) So we find that, as in example 4.1, the HK
structure can be general and that the $B$-field is self-dual up to a
shift
\be
B'= B - q/r \in H_+^2(X).
\ee
In term of the gauge theory this condition insures
that the ASD equation $\cF_+ =0$ makes sense even though $c_1$ is non-zero,
since the shifted curvature $F' =  F - 2\pi q/r$
satisfies $[\Tr F']=0$. In this case the $K3$ volume is given by
\be
V  = {v^2 \over 2r^2} - \half B'\wedge B'.
\ee

\newsection{Hyperk\"ahler geometry and moduli of moduli}

The D-brane states can be modeled in the BPS or near horizon limit in
terms of supersymmetric quantum mechanics on instanton moduli spaces.
For example the appropriate cohomology of the moduli space gives the
$1/4$ BPS states of the D-brane bound state. This quantum mechanical
system is best seen as a $\a'\ra 0$ limit of the corresponding
superconformal sigma model that we will discuss in more detail in section
7. We will now turn to a more detailed description of the properties of
these instanton moduli spaces.

\newsubsection{Hilbert schemes and moduli spaces of sheaves}

If we pick on the $K3$ surface $X$ a compatible complex structure (or
holomorphic 2-form $\eta$) and K\"ahler form $\w$, the D-brane bound
state system can be described in terms of the moduli space of suitable
coherent sheaves
\cite{dtft,morrison,harvey-moore-2}. The moduli space $\cM_v$ of simple
semi-stable (with respect to the polarization $\w$)
torsion-free sheaves $\cE$ up to equivalence with Mukai vector 
$v$ is known to be a HK manifold of dimension $4k$ with
\be
k = \half v^2 + 1.
\ee
If the vector $v$ is primitive this moduli space is smooth, compact
and simply-connected. Note that if the charge vector $v$ is null,
$\cM_v$ is a $K3$ surface itself, though not necessarily the same
$K3$. This fact is put to good use in the Mukai-Nahm Fourier
transform.

As explored by Mukai \cite{mukai,mukai-vector,mukai-map} the moduli
spaces $\cM_v$ have many beautiful properties, reviewed in for
example \cite{huybrechts,huybrechts-lehn}. First of all $\cM_v$ is a
simple compact HK space, \ie\ it is simply-connected and $h^{2,0}=1$. So
for a given complex structure the holomorphic $(2,0)$ form is unique up
to scalars. Any HK manifold can be written as a finite quotient of tori
and simple HK manifolds. 

String theorists might be surprised to learn that examples of simple
compact HK manifolds are very rare in the mathematical literature. Of
course in dimension 4 the only example is $K3$. Until very recently,
only two examples of simple HK manifolds in dimension $4k> 4$ where
known, both constructed by Beauville in 1983 \cite{beauville}. These
are respectively the Hilbert scheme $ X^{[k]}$ of zero-dimensional
schemes (read, points) of length $k$ of a $K3$ surface $X$, and the
generalized Kummer variety, which is obtained by taking the Hilbert
scheme $T^{[k+1]}$ of a four-torus $T$ and then quotienting by $T$ (so
that one obtains the usual Kummer surface representation of $K3$ for
$k=1$.) An excellent introduction to various aspects of Hilbert
schemes of complex surfaces can be found in
\cite{hilb}.

Physically these manifolds can be understood as deformations of
symmetric products of $K3$ or $T^4$ manifolds. Indeed, for any complex
surface $X$ the Hilbert scheme  is a smooth manifold that can
be obtained as a canonical desingularization 
\be
X^{[k]} \oprra^\pi \Sym^k X,
\label{proj}
\ee
of the symmetric product orbifold $\Sym^kX = X^k/S_k$. The two spaces
are closely related: \eg\ the cohomology of $X^{[k]}$ coincides with
the orbifold cohomology of $S^kX$ \cite{goettsche,gs,chea,hirzebruch}
and one expects a similar result for the elliptic genus
\cite{ell-genus}.  Note that the definition of the Hilbert scheme
requires a choice of complex structure on $X$.

If $X$ is a torus or $K3$ surface, the Hilbert scheme
has a canonical (2,0) form, essentially the pull-back of the
symmetrization of the corresponding (2,0) form on $X$. However, given a
K\"ahler class $\w$ on $X$, the Hilbert scheme does not come with a
canonical choice of K\"ahler class. In fact, one has a natural
isomorphism
\be
H^{1,1}(X^{[k]}) \cong H^{1,1}(X) \oplus \C \cdot u
\label{u}
\ee
with $u$ a class that is Poincar\'e dual to (twice) the exceptional
divisor, the inverse image of the ``small diagonal'' in $S^kX$ where at
least two points coincide. (In the orbifold cohomology of $\Sym^k X$
this cohomology class is represented by the ground state of $\Z_2$
twisted sector.) 

Generically, if $X$ contains no holomorphic curves, a K\"ahler class
$\w$ in $X$ can be lifted to any combination $\pi^* \Sym^k \w - \l u$
with $\l >0$. There is no natural value for $\l$. In the limit $\l\to 0$
the Hilbert scheme can be thought to degenerate to the symmetric
product\footnote{I thank D. Huybrechts for particularly useful
correspondence concerning this point.}. So for $\l=0$ we obtain an
orbifold metric on the Hilbert scheme in which all the fibers of the
projection (\ref{proj}) have zero volume. In this sense the symmetric
product is a point in the space of HK structures (including orbifolds)
on the Hilbert scheme.

There are various other natural constructions of simple HK manifolds.
Most relevant are the smooth ``instanton moduli spaces'' $\cM_v$ of
semi-stable torsion-free sheaves on tori or $K3$ manifolds that we
considered before. Remarkably these spaces always turn out to be HK
deformations of the Hilbert scheme. (Here there is a technical
restriction that $c_1$ should be primitive \cite{o'grady}, and thus
non-zero, that we will ignore.) More precisely, the moduli space $\cM_v$
occurs as a point in the moduli space of HK structures on the Hilbert
scheme $X^{[k]}$ for $k=\half v^2 +1$. We will see later how the HK
structure is precisely determined. (In the physics literature this fact
is sometimes stated loosely as an equivalence between $\cM_v$ and a
symmetric product.) 

Note in this context that the Hilbert scheme itself can be considered as the
special moduli space $\cM_v$ of rank one torsion-free sheaves
with $c_1=0$ and $c_2=k$ and therefore with Mukai vector
\be
v=(1,0,1-k).
\ee
In particular, as differentiable manifolds all $\cM_v$ with equal
$v^2$ (and $v$ primitive) coincide. Although some of these spaces are
birational (certainly not all of them) 
it is an open problem precisely which birational
$\cM_v$ are isomorphic as HK manifolds. 

Of course, with the usual identification between anti-self-dual
connections and holomorphic bundles, the spaces $\cM_v$ can also be
considered as instanton moduli spaces. If we just consider stable
holomorphic vector bundles these moduli spaces would in general not be
compact. By including coherent sheaves we obtain a natural, smooth
compactification that is related, but not equivalent, to the inclusion
of `point-like' instantons. For the case of rank one, where no
finite-size instantons exist, the torsion-free sheaves only represent
point-like instantons and the complicated degenerations where these
coincide.

As we discussed in \S2, these instanton moduli spaces carry natural
hyperk\"ahler metrics induced by the metric on $K3$. It might be
confusing why the choice of metric on $K3$ does not give rise to a
unique metric on the moduli space $\cM_v$. The point is precisely that
this metric has no canonical extension over the compactification. For
example, the Hilbert scheme is a natural compactification of the
symmetric product minus the diagonals. The $L^2$-metric defines a metric
on the symmetric product, but extending this metric to the Hilbert
scheme must assign a volume to the various (symplectic) blow-ups and
that introduces a free parameter $\l$. We will see in a moment that
stringy data, in the form of the self-dual NS $B$-field, will fix these
ambiguities for us.
 
There is an alternative interpretation of this issue in terms of
hyperk\"ahler quotients. If the instanton moduli space has a realization
as a HK quotient such as the ADHM construction or the
infinite-dimensional quotient discussed in \S2, the values of the HK
moment map, which is an imaginary quaternion, can be associated with the
$B$-field on $X$. Remarkably, this $B$-field can also be given a
fascinating interpretation as a deformation to a non-commutative
manifold, as demonstrated for $\R^4$ in \cite{nekrasov-schwarz,ans}. 

As we mentioned, up to recently all known constructions of simple
hyperk\"ahler manifolds were deformation equivalent to Hilbert schemes.
Only recently O'Grady has constructed a canonical desingularization of
one of the singular moduli spaces $\cM_v$ with $v$ not primitive, that
is clearly not related to a Hilbert scheme \cite{o'grady-new}. It has
$k=5$ and $b_2\geq 24$ (most likely equal to $24$) whereas the Hilbert
scheme has $b_2=23$. Note that from a physical point of view these
singular spaces are moduli spaces of multiple BPS states bound at
threshold, and therefore of great interest in studying D-brane dynamics.
Also Verbitsky has constructed examples of HK manifolds that do not seem
to be related to Hilbert schemes by considering sheaves on simple HK
manifolds \cite{verbitsky-new}.

\newsubsection{Deformations of hyperk\"ahler manifolds}

The local deformation theory of general hyperk\"ahler manifolds and the
instanton moduli spaces $\cM_v$ in particular is quite well-developed. A
good survey can be found in \cite{huybrechts}, see also
\cite{huybrechts-def,verbitsky}.

First of all, for any simple $4k$-dimensional HK manifold $Y$ there
exists a canonical quadratic form on $H^2(Y,\Z)$ constructed by
Beauville, that generalizes the intersection form for
$K3$ surfaces. It has rank $b=b_2$ and signature $(3,b-3)$. Although the
construction uses the complex structure and the associated holomorphic
$(2,0)$ form $\eta$, the final result turns out to be independent of any
choices and purely topological. Using the Hodge decomposition
\be
H^2(Y,\C) = H^{2,0}\oplus H^{1,1} \oplus H^{0,2}
\ee
Beauville's quadratic form is defined for any $w\in H^2(Y)$ as
\cite{beauville}
\be
w \cdot w = \int_Y (\eta \bar\eta)^{k-1} \Bigl( 2 \, w^{2,0} \wedge
w^{0,2} + k \, w^{1,1} \wedge w^{1,1}\Bigr).
\ee
Note that this reduces to the standard intersection form for $k=1$. The
normalization of $\eta$ can be chosen such as to make the quadratic form on
$H^2(Y,\Z)$ integral and primitive. By construction the HK 3-plane $U \subset
H^2(Y,\R)$ obtained by the period map is positive with respect to this
quadratic form. In fact, one can prove a local Torelli theorem and use
Yau's theorem to show that {\it locally} the moduli space of HK metrics
of fixed volume on a general simple HK manifold $Y$ is given by the
Grassmannian
\be
{O(3,b-3)\over O(3)\times O(b-3)},
\ee
generalizing the results for $K3$ and $T^4$ with $b=22$ and $6$
respectively. 

This result has an obvious lift to the corresponding $\cN=(4,4)$
superconformal field theory describing the sigma model on $Y$. After
including the $B$-field and the volume, and using general arguments
about the holonomy group of the Zamalodchikov metric with (4,4)
supersymmetry \cite{cecotti}, we see that the moduli space of SCFT's on
the HK manifold $Y$ is locally a quaternionic symmetric space and
therefore of the form 
\be 
{O(4,b-2)\over O(4)\times O(b-2)}. 
\ee 
It would be interesting to understand the meaning of the
underlying lattice of signature $(4,b-2)$ in this context. (We will explain the
lattice in the particular case of the Mukai moduli spaces in the next
section.)

There are however many questions about the global picture, both for the
classical geometry and the SCFT: (1) does the moduli space cover
everything (if one includes appropriate orbifolds), (2) what is the
quotient group, and (3) is the map to the Grassmannian injective, \ie\
can there be different HK structures that map to the same period in the
above Grassmannian?

For the particular case of the Hilbert scheme $X^{[k]}$ with $k>1$ of a
$K3$ surface $X$ the second cohomology group has rank 23 and can be
written as an extension of $H^2(X,\Z)\cong \G^{3,19}$ by the exceptional
divisor $u$ \cite{beauville}
\be
H^2(X^{[k]},\Z) \cong \G^{3,19}\oplus \Z \cdot u,
\qquad u^2=2-2k<0.
\label{H2}
\ee
Since the moduli spaces $\cM_v$ are all HK deformations of the Hilbert
scheme, we find quite generally the isomorphism
\be
H^2(\cM_v,\Z) \cong \G^{3,20}_k.
\ee
Of course, in the general case there is no canonical choice of the vector 
$u$, as was the case for the Hilbert scheme. But, as we discussed in \S2 all
choices of such a vector $u$ are related by a $O(3,20,\Z)$
transformation. 

We therefore find that locally the moduli space of HK structures on the 
Hilbert scheme or the Mukai spaces $\cM_v$ is given by
\be
{O(3,20)\over O(3) \times O(20)}.
\ee
Here we already recognize the form of the moduli space of attractor
$K3$ sigma models. 

If $U$ is the positive 3-plane in $H^2(X^{[k]},\R)$ determined by the HK
structure, then $U$ induces a polarization of the divisor $u$,
\be
u=u^{3,0}_L + u^{0,20}_R.
\ee
Now the condition that the Hilbert scheme degenerates to the orbifold
metric of the symmetric product can be expressed by the fact that $U$
is orthogonal to $u$, or equivalently
\be
u_L=0.
\ee
In heterotic CFT language we have a chiral vertex operator of weight
$(0,k-1)$. The appearance of this singularity is very similar to the
$A_1$ type singularities in $K3$ associated to vanishing of $-2$ curves.

\newsection{Attractor $K3$'s and instanton moduli spaces}

We will now proceed to indicate more precisely how the moduli of the
attractor $K3$ string compactification are related to the moduli of
the corresponding D-brane moduli space $\cM_v$. Hereto we make use
of the following important result. Consider a primitive Mukai vector $v\in H^*(X,\Z)$ with $v^2>0$.
Then the Mukai map (\ref{mu}) restricted to $v^\perp$ gives an isomorphism of
lattices\footnote{In \cite{o'grady} this isomorphism is only proved
for the case $c_1$ primitive and therefore non-zero, but it is
mentioned that this most likely also holds for the more general case,
at least when semi-stability implies stability.}
\cite{mukai-map,o'grady}
\be
\mu:\ v^\perp  \oprra^{\cong} H^2(\cM_v,\Z).
\ee
That is, we can canonically identify the restriction of the Mukai form
to the orthocomplement of the RR charge vector $v$ with
Beauville's ``intersection form'' on the second cohomology of the
D-brane moduli space $\cM_v$. If $v$ is null, so that $\cM_v$ itself is
a $K3$ surface, the identity is of the form $H^2(\cM_v) \cong
v^\perp/v$, an important ingredient in the Mukai-Nahm transform a.k.a.\
$T$-duality.

In fact, there is an isomorphism at the level of Hodge
structures \cite{o'grady} that can be written as
\ba
H^{2,0}(\cM_v) & \!\! \cong \!\!  & H^{2,0}(X)\nonu 
H^{1,1}(\cM_v) & \!\! \cong \!\! & \Bigl(H^0(X) \oplus
H^{1,1}(X) \oplus H^4(X)\Bigr) \cap v^\perp.
\ea
Stated otherwise, the positive two-plane $P$ in $H^2(\cM_v,\R)$ spanned
by the real and imaginary components of the holomorphic $(2,0)$ form on
$\cM_v$, coincides with the similar plane in $H^*(X,\R)$ spanned by
the holomorphic $(2,0)$ form on $X$.
So the complex structure of $\cM_v$ is directly determined
by the complex structure of the attractor $K3$ --- a fact we already
understood.

The missing ingredient is the matching of the K\"ahler forms. As we
mentioned before, there is no a priori relation between the K\"ahler
form on the $K3$ surface and the K\"ahler form on the moduli space. This
is not a surprise, since a simple counting of deformation moduli tells
us that the $B$-field on the $K3$ surface should contribute to the
determination of the K\"ahler form on the moduli space.

With our detailed description of the moduli of the string
compactification and the instanton moduli space this identification is
now straightforward. Recall that for a given Mukai or RR charge vector
$v$ the attractor moduli of the conformal field theory on $K3$ determine
a positive 3-plane $U$ in $v^\perp\otimes \R \cong \R^{3,20}$. Since we
can canonically identify $v^\perp \cong H^2(\cM_v,\Z)$ including the
Hodge structures, we see that the 3-plane $U$ can also be identified
with a three-plane in $H^2(\cM_v,\R)$. We claim $U$ is the image under
the period map of the HK structure on $\cM_v$. So we have established a
completely canonical relation between the sigma model (including the
$B$-field) on $K3$ and the classical hyperk\"ahler geometry on $\cM_v$.

Let us illustrate this discussion with the two concrete examples we
studied before.

\newsubsection{D0-D4 branes}

The D0-D4 (or D1-D5) bound state system has charge vector
$v=(r,0,-p)$, $r,p>0$ and  coprime.
In this case $v^\perp$ is spanned by $H^2(X,\Z)$ and the vector
$u=(r,0,p)$, which is primitive and satisfies
\be
u\cdot v =0, \qquad u^2=-v^2=-2rp<0.
\ee
When viewed as an element in $H^2(\cM_v)$ the vector $u$ is of type $(1,1)$
and Poincar\'e dual to the exceptional divisor in the Hilbert
scheme. So the ``period'' $\w\cdot u$ of a K\"ahler form $\w$ along $u$
determines to which extent the Hilbert scheme is deformed away from the
symmetric product.

Now, according to example (1) of \S4, the restricted deformation moduli
in this case corresponded to a general HK structure on $X$ together with
a self-dual $B$-field and volume fixed at $V=r/p - \half B^2$. One
easily sees that for $B=0$ the 3-plane $U \subset v^\perp \otimes \R
\cong \R^{3,20}$ lies entirely in $H^2(X)$. Therefore we can draw the
conclusion that for zero $B$-field (but general metric) on the $K3$
manifold, the metric on the instanton moduli space determined by the
string theory compactification is of the orbifold type that corresponds
to a symmetric product, justifying the earlier remarks in the string
literature that the D0-D4 system is described by quantum mechanics on
the symmetric product \cite{vafa}. This orbifold quantum mechanics
system should be properly understood as a limit of the corresponding
SCFT that we will discuss in the next section.

For a general non-vanishing self-dual $B$-field the corresponding
3-plane is spanned by the three vectors
\be
\Bigl(1,B+\left({\vec\w\cdot\vec\w \over \w_a\cdot B}\right) 
\w_a,p/r\Bigr)
\ee
which all have a non-vanishing inner product with the exceptional 
divisor $u$. 

\newsubsection{D2-branes}

The D2-brane system has $v=(0,q,0)$. In this case the matching of moduli
is even more involved. The lattice $v^\perp$ can be written as the
direct sum of $\G^{1,1}\cong H^0(X)\oplus H^4(X)$ and the
orthocomplement $q^\perp$ in $H^2(X)$. The restricted moduli required
$q$ to be self-dual and $q\cdot B=0$. The possible metrics on $X$ are
restricted to a $2\times19+1$ dimensional family. To obtain the full set
of $3\times 20$ deformations of $\cM_v$ we now have to include all
$2+19$ components of the $B$-field. It would be interesting to
understand this better from the point of the D2-brane gauge theory.

\newsection{The D1-D5-brane system and instanton strings}

The case of the six-dimensional strings obtained by wrapping odd
D-branes in type IIB string theory compactified on a $K3$ surface is
even more interesting, since we will now find a correspondence between on
the one hand (non-perturbative) string theory on an attractor $K3$ and
on the other hand a $c=6k$ $\cN=(4,4)$ superconformal field theory. 

\newsubsection{Type IIB on $K3$}

In the type IIB case the Ramond-Ramond fields are even-dimensional
forms and take value in $H^*(X,\R) \cong \R^{4,20}$.
Together with the string coupling constant $g_s \in \R^+$ this gives a
moduli space that is locally of the form
\be
{O(4,20)\over O(4)\times O(20)} \times \R^{4,20} \times \R_+.
\ee
Supergravity arguments indicate that the full moduli space is 
actually the symmetric space
\be
{O(5,21)\over O(5)\times O(21)}.
\ee
The full $U$-duality automorphism group is $O(\G^{5,21})$, where
$\G^{5,21}$ is the even, self-dual lattice of signature (5,21). It
contains the perturbative $T$-duality subgroup $O(\G^{4,20})$, that has an
interpretation on the level of the sigma model, but it has also extra
non-perturbative symmetries.

The occurrence of the type IIB moduli space and the lattice $\G^{5,21}$
can be explained in terms of the spectrum of six-dimensional strings.
The superstring contains besides fundamental strings also their magnetic
duals, the Neveu-Schwarz 5-branes. Furthermore there are now
odd-dimensional D-branes of dimension 1, 3, and 5. Every string or brane
can be wrapped around an even-dimensional cycle of the $K3$ manifold to
give a string in six dimensions. All in all this gives a rank $26$
lattice of strings that is isomorphic to $\G^{5,21}$. This lattice can
be considered as the direct sum of the lattice $H^*(X,\Z) \cong
\G^{4,20}$ of RR charges and an extra copy of $H^0(X,\Z)\oplus H^4(X,\Z)
\cong \G^{1,1}$ labeling the fundamental strings and NS5-branes. The
two copies of $\G^{1,1}$ are permuted by type IIB $S$-duality, that
interchanges strings with D1-branes and NS 5-branes with D5-branes
(D3-branes are self-dual). In fact, it is an elegant result that the
full $U$-duality group $O(\G^{5,21})$ is generated by the $\Z_2$
$S$-duality together with the $T$-duality group $O(\G^{4,20})$
\cite{aspinwall}.

There are again formulas that express how the standard sigma moduli of
$K3$, combined into the quaternionic 2-form $\bb$, 
together with the Ramond-Ramond gauge fields 
\be
\AR=(\th,\BR,G) \in H^*(X,\R)
\ee
and the string coupling $g_s\in\R_+$, determine a positive 5-plane $Z$ in 
$\R^{5,21}$ . Writing the charge lattice in terms of
Ramond and Neveu-Schwarz charges as $\G^{5,21}=\G_{R}^{4,20}\oplus
\G_{NS}^{1,1}$, the 5-plane is spanned by the vectors \cite{aspinwall}
\be
\bigl(\exp\bb,0,\AR\cdot \exp\bb\bigr),\quad
\bigl(\AR,1,1/g_s\bigr).
\ee

\newsubsection{D-branes and their moduli}

A six-dimensional string with charge vector $v\in \G^{5,21}$ can be
described in the decoupling or near-horizon limit as a two-dimensional
$\cN=(4,4)$ superconformal field theory. All primitive vectors $v$ of
equal length are equivalent under $U$-duality, and we can choose the
vector $v$ to lie in the RR lattice $H^*(X,\Z) \cong \G^{4,20}$. In that
case the SCFT should be identified with the sigma model on the Mukai
moduli space $\cM_v$. The attractor moduli of the type IIB $K3$
compactifications should now be matched with the HK metric and $B$-field
of $\cM_v$. 

We see directly that the local structure of the moduli spaces coincides.
Picking a charge vector $v\in \G^{5,21}$ with $v^2 = 2k-2>0$ leads to an
attractor moduli space that describes positive 4-planes $W$ in
$v^\perp\otimes\R \cong \R^{4,21}$ and this is given by the Grassmannian
\be
{O(4,21)\over O(4)\times O(21)}.
\label{4,21}
\ee
There is a stabiliser $U$-duality subgroup $O(\G^{4,21}_k)$ with
\be
\G^{4,21}_k \cong \G^{3,20}_k \oplus \G^{1,1}.
\ee
If we have only D-brane charges, \ie\ if the (primitive)
charge vector $v$ is of the form
\be
v=(Q_5,Q_3,-Q_1) \in H^*(X,\Z),
\ee
the attractor moduli are described completely analogously to example
4.2. The fixed point conditions read (with $\zz \in \H$ a quaternion to be
determined)
\be
\Re(\zz\bb) = v,\qquad \bb \cdot \AR = 0.
\ee
These equations have an obvious solution.
First of all the 4-plane $W\subset H^*(X,\R)$ determined by the 
``quaternionic'' K\"ahler form $\bb$ should contain the vector $v$. 
This condition is equivalent to the fact that the
$K3$ sigma model is attractive for the D-brane charge vector $v$.

Secondly, the RR fields $\AR$ should satisfy 
\be
v \cdot \AR = Q_1 \, \theta + Q_3 \cdot \BR - Q_5\, G = 0.
\label{flux}
\ee
So the total flux of RR fields through the collection of D-branes should
vanish. The RR fields therefore lie in $v^\perp\otimes \R \cong
\R^{3,20}$ (modulo shifts). In this case the string coupling constant $g_s$
is not fixed. This description corresponds to the decomposition of
(\ref{4,21}) as
\be
{O(3,20)\over O(3)\times O(20)} \times \R^{3,20} \times \R_+
\ee

In this form we immediately recognize the moduli of the CFT on $\cM_v$.
We already saw in the previous section how the sigma model moduli of
$K3$ relate to the HK structure on $\cM_v$ (the first factor). So this
identification works just as well in the CFT. In particular, in the
absense of NS $B$-fields, the HK metric is of the symmetric product
orbifold form. Now the IIB string RR fields $\AR$ can be related to the
sigma model $B$-field, using the identification $H^2(\cM_v) \cong
v^\perp$ (the second factor). Finally, the inverse string coupling is
identified with the volume or equivalently $\a'$ of the sigma model
(the last factor). 

These identifications can also be understood physically from the D-brane
gauge theory point of view as we discussed in \S2. We have already
remarked that the string coupling plays the role of $\a'$ for the $c=6k$
CFT, so that the weakly coupled IIB string regime coincides with sigma
model perturbation theory of the instanton string. Note that in the
absense of D3-branes and with $B$ and $\AR$ set to zero the volume of
the $K3$ surface is fixed to be $Q_1/Q_5$.

The fact that the RR field $\BR$ becomes a sigma model $B$-field on the
instanton string is not surprising, since that string is essentially the
D-string and $S$-duality tells us that the D-string couples to $\BR$
just as the fundamental string couples to $B$. Also, from the D5-brane
action we have derived that the RR gauge fields produce two-form
$B$-fields for the instanton string. More precisely, the RR background field
$\BR$ induces the
forms $\cB_I$ of (\ref{BR}) which can be identified with the usual
cohomology classes of Donaldson theory, the descendents of $\Tr F^2$.
Similarly the $\th$-angle produces an extra field of the form
$\cB_0$ (\ref{BF}), a descendent of $\Tr F^3$. Furthermore the Mukai map
from the RR charge lattice to the second cohomology of the instanton
moduli space is exactly induced by the D5-brane couplings.

We now want to understand in string theory terms the condition
(\ref{flux}) that the total RR flux through the D-branes should be zero,
which forces a non-zero coupling to the 4-form field $C$. One way to
explain this constraint is that in the reduction from the
six-dimensional SYM theory to the two-dimensional sigma model we also
obtain a two-dimensional Yang-Mills field with curvature $f$. In general
there can be a FI term coupling to the flux of the $U(1)$ curvature $\Tr
f$. Working through the reduction we see that this FI term is induced by
the RR background fields as
\be
%\int_X \Bigl( Q_1 \, \theta + Q_3 \cdot \BR - Q_5\,G\Bigr)\cdot 
(v \cdot \AR) \int_\Sigma\Tr f
\ee
So in order to avoid this term we have to choose the combination of $\AR$
such that the total flux vanishes.

\newsubsection{Global issues and $u$-duality}

Let us finally comment on some of the global issues of the various moduli spaces
that we have discussed so far. Let us begin with the map between the
$K3$ CFT moduli and the classical HK geometry of the instanton moduli
space $\cM_v$. Starting from the $T$-duality of the $K3$ sigma model, one
derives that the global moduli space of the perturbative D-brane system,
that is obtained in the $g_s \ra 0$ limit, is given by the Narain space
\be
O(\G^{3,20}_k)/ O(3,20)\backslash O(3)\times O(20).
\label{global}
\ee
We have argued that this moduli space should correspond to the space of
HK metrics of fixed (in fact, very large) volume on the (unmarked) Mukai
moduli space $\cM_v$. The space (\ref{global}) is the classical HK
period domain and it is known that the moduli space of HK structures
maps finitely into this period domain. It is further believed that this
map is generically injective \cite{daniel}. String theory clearly seems
to suggests that the map is surjective if one includes singular
orbifold metrics. (We will momentarily return to the injectivity.)

Similarly there is a statement for the D-brane conformal field theory,
where $U$-duality now implies that the global form of the space of
deformations is
\be
O(\G^{4,21}_k)/ O(4,21)\backslash O(4)\times O(21). 
\label{moduli}
\ee
Here the $U$-duality stabilizer group $O(v^\perp)=O(4,21,\Z)$ has an
interpretation as a $T$-duality of the $\cN=(4,4)$ SCFT. 

When we interpret this sigma model as a DLCQ of the six-dimensional
little string theory on $K3 \times \R^{1,1}$, the four extra moduli
compared to the usual 80 $K3$ moduli are interpreted as a (quaternionic)
string coupling constant \cite{5d,witten-higgs}. (In matrix theory these
same couplings correspond to the components $C_{\m\n-}$ of the
background 3-form gauge field of 11-dimensional M-theory \cite{abs}.)
The group $O(4,21,\Z)$ interchanges this string coupling with the
geometric $K3$ moduli, and therefore can be considered as a ``little
$u$-duality'' of the six-dimensional string theory. It is nice to see
in this concrete example how intrinsic non-geometric objects like the
string coupling are treated on equal footing with the geometric ones. It
would be very interesting to find an ``m-theory'' interpretation of this
duality \cite{m&m}.

There is however a more interesting point with possible mathematical
implications. All D-branes with Mukai vectors of equal length are
related by dualities. Therefore the corresponding SCFT's should be
isomorphic, after a possible shift in their moduli. We know that this
particular component of the moduli space of $c=6k$ $\cN=(4,4)$ SCFT's
takes the form (\ref{moduli}) and we can find the particular point a
certain D-brane configuration is mapped to by the period map that we
discussed in detail.

At the level of classical HK structures the global structure of the
moduli space is actually not known. Taking the classical, large volume
limit of the SCFT moduli space, string theory seems to arrive at the
(\ref{global}) as the moduli space of HK structures of fixed volume
--- a very strong statement, indicating a global Torelli theorem for
the Hilbert scheme. In general it is not known if every two Mukai
spaces with the same period point are isomorphic as HK
manifolds. However, it is known that these spaces are birationally
equivalent \cite{huybrechts}.  Furthermore, as soon as one deforms the
complex structure away from the point where the spaces have an
interpretation as moduli spaces of sheaves on $K3$, the deformed
spaces become isomorphic as HK manifolds. Therefore the moduli space
of HK structures seems to be a priori non-Hausdorff with non-separated
points.

Can we have a non-Hausdorff moduli space of SCFT's? The general idea
of conformal perturbation theory, where the neighbourhood of a CFT is
parametrized by the exactly marginal operators, combined with the $\cN
= (4,4)$ supersymmetry suggests that it is not the case. (See however
\cite{finite}, although that example is of a somewhat different nature
because of the time-like compactification.). It seems difficult to
give a completely rigorous argument that conformal perturbation theory
works in this case and that the moduli space of $\cN=(4,4)$ SCFT is
Hausdorff, although everything in string theory and supergravity
indicates that it is. In the work of Aspinwall and Morrison on $K3$
compactifications \cite{aspinwall-morrison}, which underlies much what
has been said here, it was an input that the moduli space was
Hausdorff. Therefore it is not a surprise that it also comes out in
the form of the nice arithmetic quotients (\ref{global}) and
(\ref{moduli}). Of course there are many examples of equivalent string
compactifications on manifolds that are only birational \cite{agm},
and therefore an other possibility is the CFT moduli space is better
behaved than the classical geometries. This point clearly deserves
further study.

\vspace{8mm}

{\noindent \bf Acknowledgements}

\vspace{2mm}

I would like to acknowledge useful discussions with S. Agnihotri,
J. de Boer, L.  G\"ottsche, B. Fantechi, C. Hofman, M. Kontsevich,
E. Looijenga, E. Martinec, G. Moore, J.-S.  Park, E. Verlinde,
H. Verlinde, and in particular D. Huybrechts.

\renewcommand{\Large}{\normalsize}

\end{document}